\documentclass[journal]{IEEEtran}
\usepackage{amsmath, amsthm, amsfonts, amssymb}
\usepackage{graphicx}
\usepackage{multirow}
\usepackage{enumerate}
\usepackage{subfigure}
\usepackage[table,xcdraw]{xcolor}
\usepackage{amsmath}
\usepackage{diagbox}
\usepackage{bm}
\usepackage{algorithmic,algorithm}
\usepackage{etoolbox}
\usepackage{color,xcolor}

\newtheorem{theorem}{\bf Theorem}

\newtheorem{lemma}[theorem]{\bf Lemma}

\newcounter{definition}

\long\def\symbolfootnote[#1]#2{\begingroup
\def\thefootnote{\fnsymbol{footnote}}\footnote[#1]{#2}\endgroup}

\begin{document}

\title{Dynamic Control for Random Access in Deadline-Constrained Broadcasting}

\author{Aoyu Gong, Lei Deng, Fang Liu and Yijin Zhang
\thanks{This work was supported in part by the National Natural Science Foundation of China under Grants 62071236, 61902256, and in part by the Fundamental Research Funds for the Central Universities of China (No.~30920021127). (\emph{Corresponding author: Yijin Zhang}.)}
\thanks{A. Gong and Y. Zhang are with the School of Electronic and Optical Engineering, Nanjing University of Science and Technology, Nanjing 210094, China.
E-mail: \{gongaoyu; yijin.zhang\}@gmail.com.}
\thanks{L. Deng is with the College of Electronics and Information Engineering, Shenzhen University, Shenzhen 518061, China. E-mail: ldeng@szu.edu.cn.}
\thanks{F. Liu is with the Department of Information Engineering, The Chinese University of Hong Kong, Shatin, N. T., Hong Kong. E-mail: lf015@ie.cuhk.edu.hk.}}

\maketitle

\begin{abstract}
This paper considers random access in deadline-constrained broadcasting with frame-synchronized traffic.
To enhance the maximum achievable timely delivery ratio (TDR), we define a dynamic control scheme that allows each active node to determine the transmission probability with certainty based on the current delivery urgency and the knowledge of current contention intensity.
For an idealized environment where the contention intensity is completely known, we develop an analytical framework based on the theory of Markov Decision Process (MDP), which leads to an optimal scheme by applying backward induction.
For a realistic environment where the contention intensity is incompletely known, we develop a framework using Partially Observable Markov Decision Process (POMDP), which can in theory be solved.
We show that for both environments, there exists an optimal scheme that is optimal over all types of policies.
To overcome the infeasibility in obtaining an optimal or near-optimal scheme from the POMDP framework, we investigate the behaviors of the optimal scheme for two extreme cases in the MDP framework, and leverage intuition gained from these behaviors to propose a heuristic scheme for the realistic environment with TDR close to the maximum achievable TDR in the idealized environment.
In addition, we propose an approximation on the knowledge of contention intensity to further simplify this heuristic scheme.
Numerical results with respect to a wide range of configurations are provided to validate our study.
\end{abstract}


\IEEEdisplaynontitleabstractindextext
\IEEEpeerreviewmaketitle

\section{Introduction} \label{sec:introduction}
\IEEEPARstart{B}{roadcasting} is a fundamental operation in wireless networks.
With the explosive growth of ultrareliable low-latency services for Internet of things~\cite{2019urllc,2018Gao,2019Luvi}, such as multimedia sharing in sensor networks, safety message dissemination in vehicular networks and industrial control in factory automation systems, deadline-constrained broadcasting has been becoming a research focus in recent years.
For such broadcasting, each packet needs to be transmitted within a strict delivery deadline since its arrival, and will be discarded if the deadline expires.
Hence, \emph{timely delivery ratio} (TDR), defined as the probability that a broadcast packet is successfully delivered to an arbitrary intended receiver within the given delivery deadline, is considered as a critical metric to evaluate the performance of such broadcasting.

A canonical deadline-constrained broadcasting scenario is that an uncertain set of nodes with new or backlogged packets attempt to transmit at approximately the same time, so that random access mechanisms are needed to support efficient channel sharing and careful design of access parameters is needed to maximize the TDR.

Many recent literatures have been dedicated to this issue.
Under saturated traffic, Bae~\cite{2013CRN, 2013RA} obtained the optimal slotted-ALOHA for broadcasting single-slot packets and optimal $p$-persistent CSMA for broadcasting multi-slot packets, respectively.
Under a discrete-time Geo/Geo/1 queue model, Bae~\cite{2015Queue} obtained the optimal slotted-ALOHA for broadcasting single-slot packets.
Under frame-synchronized traffic, Campolo~\emph{et al.}~\cite{2011WAVE} proposed an analytical model for using IEEE 802.11p CSMA/CA to broadcast multi-slot packets, which can be used to obtain the optimal contention window size.
However,~\cite{2013CRN, 2013RA,2015Queue} adopted a static transmission probability and~\cite{2011WAVE} adopted a static contention window size, thus inevitably limiting the maximum achievable TDR.
Other studies on deadline-constrained random access include~\cite{2018asymAloha,2019mprAloha} for uplink to a common receiver and~\cite{2020TTS} for unicast in ad hoc networks, which all still restrict their attentions to static access parameters.

As such, to enhance the maximum achievable TDR of random access in deadline-constrained broadcasting, it is strongly required to develop a dynamic control scheme that allows each node to adjust its access parameters according to the knowledge of current delivery urgencies and the knowledge of current contention intensity.
Unfortunately, due to random traffic or limited capability on observing the channel status, each node \emph{cannot} obtain a complete knowledge of the current contention intensity in practice, which renders such design a challenging task.
So, each node has to estimate the current contention intensity using the information obtained from the observed channel status.
A great amount of work has gone into studying such information that can be obtained~\cite{Segall1976,1987Bayesbroadcast,2004fine,2004Dynamic802.11,2013FASA} under various models and protocols.
Our work follows the same direction of~\cite{1987Bayesbroadcast,2004fine} to keep an A Posteriori Probability (APP) distribution for the current contention intensity given all past observations and access settings, which is a sufficient statistic for the optimal design~\cite{1971POMDP}, but needs to additionally take into account the impact of delivery urgencies.
To our best knowledge, this is the first time to study dynamic control for deadline-constrained random access.

Furthermore, it is naturally desirable for this dynamic control to strike a balance between the chance to gain an instantaneous successful transmission and the chance to gain a future successful transmission within the given deadline, which requires reasoning about future sequences of access parameters and observations.
So, the dynamic control design under this objective is more challenging than that for maximizing the instantaneous throughput of random access~\cite{1987Bayesbroadcast,2004Dynamic802.11,2013FASA}, which is only ``one-step look-ahead''.
By seeing access parameters as actions, in this paper we apply the theories of Markov Decision Process (MDP) and Partially Observable MDP (POMDP) to obtain optimal control schemes.
Although the idea of using MDP and POMDP in the context of random access control is not new~\cite{2004fine,2019pcsma,2018energy}, to our best knowledge, this is the first work to apply them in deadline-constrained broadcasting.
In addition, as solving POMDP is in general computationally prohibitive, it is important to develop a simple control scheme with little performance loss.

In this paper, we focus on deadline-constrained broadcasting under frame-synchronized traffic.
Such a traffic pattern can capture a number of scenarios in machine-to-machine communications~\cite{2019urllc,2011WAVE,2018asymAloha,2020TTS,2019traffic} where each node has periodic-\emph{i.i.d.} packet arrivals.
The contributions of our work are as follows.
\begin{enumerate}
  \item We generalize slotted-ALOHA to define a dynamic control scheme, i.e., a deterministic Markovian policy, which allows each active node to determine the current transmission probability with certainty based on its current delivery urgency and the knowledge of current contention intensity.
  \item For an idealized environment where the contention intensity is completely known, we develop an analytical framework based on the theory of MDP, which leads to an optimal control scheme by applying the backward induction algorithm.
      We further show it is indeed optimal over all types of policies for this environment.
  \item For a realistic environment where the contention intensity is incompletely known, we develop an analytical framework based on the theory of POMDP, which can in theory lead to an optimal control scheme by backward induction.
      We also show it is indeed optimal over all types of policies for this environment.
  \item To overcome the infeasibility in obtaining an optimal or near-optimal control scheme from the POMDP framework, we investigate the behaviors of the optimal control scheme for two extreme cases in the idealized environment, and use these behaviors as clues to design a simple heuristic control scheme for the realistic environment with TDR close to the maximum achievable TDR in the idealized environment.
     In addition, we propose an approximation on the knowledge of contention intensity to further simplify this heuristic scheme.
  %\item We modify the above frameworks to incorporate random fading effect.
  % In particular, built on the underpinnings above and an observation-reduction method, we propose a heuristic scheme for the realistic-knowledge environment with random channel errors.
\end{enumerate}
Note that although the MDP framework for the idealized environment has limited applicability as the contention intensity \emph{cannot} be completely known in practice, it will serve to provide an upper bound on the maximum achievable TDR in the realistic environment, and serve to provide clues to design a heuristic scheme for the realistic environment.

The remainder of this paper is organized as follows.
The system model is specified in Section~\ref{sec:system}, and a dynamic control scheme is defined in Section~\ref{sec:strategy}.
The idealized and realistic environments are studied in Sections~\ref{sec:mdpframework} and~\ref{sec:pomdpframework}, respectively.
A simple heuristic control scheme for the realistic environment is proposed in Section~\ref{sec:heuristicstrategy}.
Numerical results with respect to a wide range of configurations are provided in Section~\ref{sec:numerical}.
Conclusions are given in Section~\ref{Conclusion}.

\section{System Model and Dynamic Control Scheme} \label{sec:systemstrategy}
\subsection{System model} \label{sec:system}
Consider a globally synchronized wireless network where a finite number, $N\geq 2$, of nodes are within the communication range of each other.
The global time axis is divided into frames, each of which consists of a finite number, $D \geq 1$, of time slots of equal duration, indexed by $t \in \mathcal{T} \triangleq \{1,2,\dots,D\}$.
To broadcast the freshest information, at the beginning of each frame, each node independently generates a packet to be transmitted with probability $\lambda\in (0,1]$.
We further assume that every packet has a strict delivery deadline $D$ slots, i.e., a packet generated at the beginning of a frame will be discarded at the end of this frame.
%For ease of presentation, we restrict ourselves to single-slot packets, and it is straightforward to extend our work to consider multi-slot packets.
A broadcasting scenario for collaborative target
detection with the above assumptions can be found in Fig.~1.

\begin{figure}[!ht]
	\centering
	\includegraphics[width=3.5in]{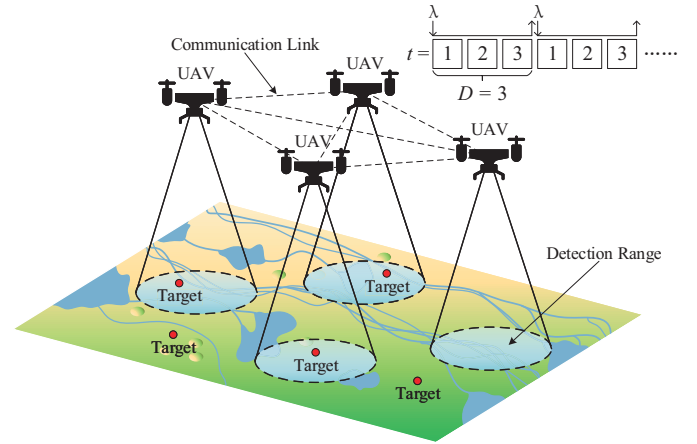}
	\caption{A deadline-constrained broadcasting scenario for collaborative target detection}
\end{figure}

By considering random channel errors due to wireless fading effect, we assume that a packet sent from a node is successfully received by an arbitrary other node with the
probability $\sigma \in (0, 1]$ if it does not collide with other packets, and otherwise is certainly unsuccessfully received by any other node.
%This assumption is quite accurate when every packet is well protected by error correction codes.
Due to the broadcast nature, we assume that every packet is neither acknowledged nor retransmitted.
Then, at the beginning of slot $t$ of a frame, a node is called an \emph{active node} if it generated a packet at the beginning of the frame and has not transmitted before slot $t$, but is called an \emph{inactive node} otherwise.
Each active node follows a common control scheme for random access, which will be defined in Section~\ref{sec:strategy}, to generate transmission probabilities at the beginnings of different slots.

A slot is said to be in the \emph{idle} channel status if no packet is being transmitted, and in the \emph{busy} status otherwise.
%At the end of a slot, we assume that each node that is not transmitting in this slot is able to be aware of the channel status of this slot.
At the end of a slot, we assume that each node is able to be aware of the channel status of this slot.
%but is not able to distinguish the number of packets involved in a failure due to the limited detection capability.

The values of $N$, $D$, $\lambda$ and $\sigma$ are all assumed to be completely known in advance to each node.

\subsection{Dynamic Control for Random Access} \label{sec:strategy}
Due to frame-synchronous traffic, at the beginning of an arbitrary slot with at least one active node, we know each active node has the same delivery urgency.
To take into account the joint impact of delivery urgency and the knowledge of contention intensity on determining transmission probabilities, a dynamic control scheme for random access in deadline-constrained broadcasting, which can be seen as a generalization of slotted-ALOHA, is formally defined as follows.

Consider an arbitrary frame.
Let the random variable $n_t$ taking values in $\mathcal{N}\triangleq\{0,1,\ldots,N-1\}$ denote the actual number of other active nodes in the view of an arbitrary node at the beginning of slot $t \in \mathcal{T}$.
At the beginning of an arbitrary slot $t \in \mathcal{T}$ with at least one active node, we assume that each active node has the same observation history (for estimating the actual value of $n_t$) from the environment, and we require each active node to adopt the same transmission probability.
Thus, each active node has the same knowledge of the actual value of $n_t$ based on all past observations from the environment and all past transmission probabilities.
Such a knowledge can be summarized by a probability vector $\mathbf{b}_t \triangleq \big(b_t(0), b_t(1),\ldots, b_t(N-1) \big)$, called the \emph{activity belief}, where $b_t(n)$ is the conditional probability (given all past observations from the environment and all past transmission probabilities) that $n_t=n$.
Let $\mathcal{B}_t$ denote the set of all possible values of $\mathbf{b}_t$ in $[0,1]^N$ such that $\sum_{n=0}^{N-1} b_t(n)=1$.
Hence, at the beginning of every slot $t\in \mathcal{T}$ with at least one active node, we require each active node to use the values of $t$ and $\mathbf{b}_t$ for determining the value of transmission probability $p_t$ by a transmission function $\pi_t: \mathcal{B}_t \rightarrow [0,1]$.
An example of the working procedure for the case of $N = 8$, $D = 6$ is illustrated in Fig.~1.

\begin{figure}[!ht]
	\centering
	\includegraphics[width=3.4in]{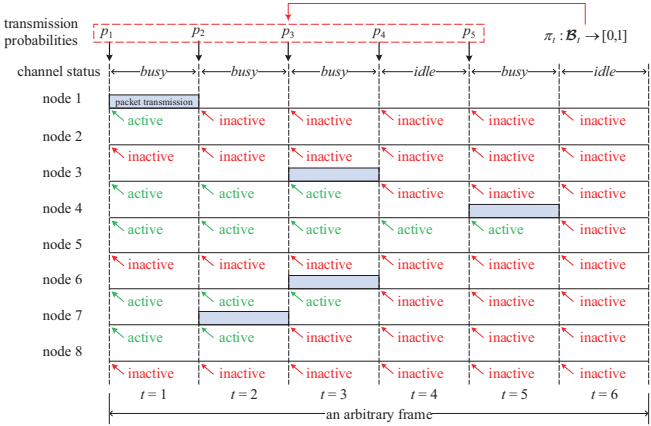}
	\caption{An example of the working procedure for $N=8$, $D=6$.}
\end{figure}

We further consider two different environments for active nodes to obtain the value of the activity belief $\mathbf{b}_t$.
\begin{enumerate}
    \item \underline{\emph{Idealized environment:}} at the beginning of every slot $t\in \mathcal{T}$ with at least one active node, each active node always has a complete knowledge of the value of $n_t$, i.e., $\mathbf{b}_t=(0,\ldots,0, b_t(n)=1,0,\ldots,0)$ if $n_t=n$ actually.
        Hence, the transmission function $\pi_t$ in this environment can be simply written as a function $\widehat{\pi}_t: \mathcal{N} \rightarrow [0,1]$.

    \item \underline{\emph{Realistic environment:}} at the beginning of each slot $t\in \mathcal{T}$ with at least one active node, each active node is able to obtain the value of $\mathbf{b}_t$ only based on the characteristic of packet arrivals, all past channel statuses (idle or busy) and all past transmission probabilities, and thus has an incomplete knowledge of the value of $n_t$.

\end{enumerate}
Obviously, the idealized environment is infeasible to implement due to the difficulty in determining the initial actual number of other active nodes and determining the number of nodes involved in each busy slot, whereas the realistic environment can be easily implemented without imposing extra overhead and hardware cost.

The objective of subsequent sections is to seek optimal control schemes, i.e., design $\widehat{\pi}_t$ and $\pi_t$ sequentially in each slot so that the TDR is maximized, for the idealized and realistic environments, respectively.
It will be shown in Section~\ref{sec:mdpsolution} that the mapping of $\widehat{\pi}_t$ can lead to an optimal control scheme over all possible schemes for the idealized environment, and will be shown in Section~\ref{sec:pomdpsolution} that the mapping of $\pi_t$ can lead to an optimal control scheme over all possible schemes for the realistic environment.

\section{MDP Framework for the Idealized Environment} \label{sec:mdpframework}
In this section, we formulate the random access control problem in the idealized environment as an MDP, use the expected total reward of this MDP to evaluate the TDR, and then obtain an optimal control scheme that maximizes the TDR.

\subsection{MDP Formulation} \label{sec:mdpformulation}
For an arbitrary frame, consider an arbitrary node as the tagged node, and let the random variable $q_t$ taking values from $\{0\,(\text{inactive}),1\,(\text{active})\}$ denote the status of the tagged node at the beginning of slot $t$.
From the dynamic control scheme specified in Section~\ref{sec:strategy}, we see that each node becomes inactive at the beginning of slot $t+1$ with the transmission probability $p_t$ if it is active at the beginning of slot $t$, and will be always inactive if it is inactive at the beginning of slot $t$.
This implies that the probability of moving to the next state in the state process $(q_t,n_t)_{t \in \mathcal{T}}$ depends only on the current state.
Thus, $(q_t,n_t)_{t \in \mathcal{T}}$ can be viewed as a discrete-time finite-horizon, finite-state Markov chain.

Based on the Markov chain $(q_t, n_t)_{t \in \mathcal{T}}$, we present an MDP formulation by introducing the following definitions.

\begin{enumerate}
\item \underline{\emph{Actions:}}
    At the beginning of each slot $t \in \mathcal{T}$ with $q_t=1$, the action performed by the tagged node (and the other active nodes) is the chosen transmission probability $p_t$ taking values in the action space $[0,1]$.
    Note that the tagged node performs no action when $q_t = 0$.

\item \underline{\emph{State Transition Function:}}
    As the tagged node will never transmit since slot $t$ if $q_t=0$, we only concern about the state transition function when $q_t=1$.
    The state transition function $\beta_t \big((q',n'),(1,n),p\big)$ is defined as the transition probability of moving from the state $(q_t,n_{t})=(1,n)$ to the state $(q_{t+1},n_{t+1})=(q',n')$ when each active node at the beginning of slot $t$ adopts the transmission probability $p_t=p$.
    So, we have
    \begin{align}
    & \beta_t\big((q',n'),(1,n),p\big) \notag \\
    & \triangleq \text{Pr}\big((q_{t+1},n_{t+1})=(q',n')|(q_t,n_{t})=(1,n),p_t=p\big) \notag \\
    & =
    \begin{cases}
        \binom{n}{n-n'} p^{n-n'+1-q'}(1-p)^{n'+q'},  &\text{if}\,\, n'\leq n,\\
        0, &\text{otherwise},
    \end{cases} \label{trans}
    \end{align}
   for each $t\in \mathcal{T}\setminus \{D\}$, each $q' \in \{0,1\}$, each $n,n' \in \mathcal{N}$ and each $p\in [0,1]$.

\item \underline{\emph{Rewards:}}
     The reward gained at slot $t$ is defined as the average number of packets of the tagged node transmitted successfully to an arbitrary other node at slot $t$.
     As there is no reward at slot $t$ when $q_t=0$, we only focus on the cases when $q_t= 1$.
     Let $r_t\big((1,n),p\big)$ denote the reward at slot $t$ for the state $(q_t,n_t) =(1,n)$ when each active node at the beginning of slot $t$ adopts the transmission probability $p_t=p$.
     So, we have
     \begin{equation} \label{reward}
        r_t\big((1,n),p\big) = \sigma p(1-p)^{n},
     \end{equation}
     for each $t\in \mathcal{T}$, each $n \in \mathcal{N}$ and each $p \in  [0,1]$.

\item \underline{\emph{Policies:}}
A deterministic Markovian policy $\widehat{\bm{\pi}}$ is defined by a sequence of transmission functions (i.e., decision rules) for the idealized environment:
\begin{equation}
\widehat{\bm{\pi}} \triangleq (\widehat{\pi}_1,\widehat{\pi}_2,\ldots,\widehat{\pi}_D),  \quad \text{where } \widehat{\pi}_t: \mathcal{N} \rightarrow   [0,1]. \notag
\end{equation}
Let $\widehat{\bm{\Pi}}^{\text{MD}}$ denote the set of all possible such polices.
Obviously, a dynamic control scheme for the idealized environment as described in Section~\ref{sec:strategy} is essentially a deterministic Markovian policy here.
\end{enumerate}

Let $R^{\widehat{\bm{\pi}}}(1,n)$ denote the expected total reward from slot $1$ to slot $D$ when $q_1=1$, $n_1 =n$ and the policy $\widehat{\bm{\pi}}$ is used, which can be defined by
\begin{align}
    & R^{\widehat{\bm{\pi}}}(1,n) \notag \\
    & \triangleq \mathbb{E}^{\widehat{\bm{\pi}}} \Big\{ \sum_{t=1, q_t=1}^D  r_t \big( (1,n_t),\widehat{\pi}_t(n_t) \big) | q_1 =1, n_1=n \Big\}, \notag
\end{align}
where $\mathbb{E}^{\widehat{\bm{\pi}}}$ represents the conditional expectation given that policy $\widehat{\bm{\pi}}$ is employed.
Then, the TDR under the policy $\widehat{\bm{\pi}}$ can be computed by
\begin{equation}
\text{TDR}^{\widehat{\bm{\pi}}}=\sum_{n \in \mathcal{N}} \binom {N-1}{n} \lambda^n(1-\lambda)^{N-1-n}   R^{\widehat{\bm{\pi}}}(1,n). \notag
\end{equation}

\subsection{MDP solution} \label{sec:mdpsolution}
Due to the finite horizon, finite state space, compact action space, bounded rewards, continuous rewards with respect to $p$ and continuous state transition function with respect to $p$ in our MDP formulation,
\cite[Prop.~4.4.3, Ch.~4]{pm2014} indicates that for maximizing $\text{TDR}^{\widehat{\bm{\pi}}}$, there exists a $\widehat{\bm{\pi}} \in \widehat{\bm{\Pi}}^{\text{MD}}$, which is indeed optimal over all random and deterministic, history-dependent and Markovian policies.
This property also justifies the transmission function and design goal for the idealized environment considered in Section~\ref{sec:strategy}.
Hence, we aim to seek
\begin{equation}
  \widehat{ \bm{\pi}}^* \in \mathop{\arg\max}_{\widehat{\bm{\pi}} \in \widehat{\bm{\Pi}}^{\text{MD}}} \text{TDR}^{\widehat{\bm{\pi}}}. \notag
\end{equation}

Let $U^*_t(1,n)$ denote the value function corresponding to the maximum total expected reward from slot $t$ to slot $D$ when $q_t=1$ and $n_t = n$. Averaging over all possible next states with $q_{t+1}=1$, we arrive at the following Bellman's equation:
\begin{equation}\label{BellmanI}
\begin{aligned}
     & U^*_D(1,n) = \mathop{\max}_{{p \in  [0,1]}} r_D\big((1,n),p\big),  \quad \forall n \in \mathcal{N},\\
     & U^*_t(1,n) = \mathop{\max}_{{p \in  [0,1]}} r_t\big((1,n),p\big) \\
     & \ \quad + \sum_{n' \in \mathcal{N}} \beta_t\big((1,n'),(1,n),p\big) U^*_{t+1}(1,n'), \quad \forall n \in \mathcal{N},
\end{aligned}
\end{equation}
for each $t\in \mathcal{T}\setminus \{D\}$.
Then, applying the backward induction algorithm to get a solution to Eq.~\eqref{BellmanI} involves finding global maximizers of a series of real-coefficient univariate polynomials defined on $[0,1]$, and can formally lead to $\widehat{\bm{\pi}}^*$.

\section{POMDP Framework for the Realistic Environment} \label{sec:pomdpframework}
In this section, we formulate the random access control problem in the realistic environment as a POMDP, use the expected total reward of this POMDP to evaluate the TDR, and then discuss how to obtain an optimal or near-optimal control scheme.

\subsection{POMDP Formulation} \label{sec:pomdpformulation}
Based on the Markov chain $(q_t, n_t)_{t \in \mathcal{T}}$ specified in Section~\ref{sec:mdpframework} and the activity belief $\mathbf{b}_t$ for each $t \in \mathcal{T}$ with $q_t=1$ specified in Section~\ref{sec:strategy}, we present a POMDP formulation by introducing the following definitions.

\begin{enumerate}
    \item \underline{\emph{Actions, State Transition Function, Rewards:}}
    The definitions of these elements are the same as in Section~\ref{sec:mdpframework}.

    \item \underline{\emph{Observations and Observation Function:}}
     The tagged node at the beginning of slot $t+1$ can obtain an observation on the channel status of slot $t$, denoted by $o_t$.
    When $q_{t+1}=0$, the tagged node will never transmit since slot $t+1$ and $o_t$ will thus be useless.
    Hence, we only consider $o_t$ when $q_{t+1}=1$ taking values from the observation space $\mathcal{O} \triangleq \{ 0\,(\text{idle}),1\,(\text{busy}) \}$.
    Further, the observation function $\omega_t\big(o,(1,n),(1,n')\big)$ is defined as the probability that the tagged node at the beginning of slot $t+1$ obtains the observation $o_{t}=o$ if the state $(q_{t},n_{t})=(1,n)$ and the state $(q_{t+1},n_{t+1})=(1,n')$. So, we have
\begin{align}
& \omega_t \big(o,(1,n),(1,n')\big) \notag \\
& \triangleq \text{Pr} \big(o_{t}=o|(q_{t},n_{t})=(1,n),(q_{t+1},n_{t+1})=(1,n')\big) \notag \\
&  =
\begin{cases}
1,  &\text{if}\,\, o=0,n=n', \\
1,  &\text{if}\,\, o=1,n-n'\geq 1, \\
0, &\text{otherwise}, \notag
\end{cases}
\end{align}
    for each $t\in \mathcal{T}$, each $o \in \mathcal{O}$ and each $n, n' \in \mathcal{N}$.

    \item \underline{\emph{Bayesian update of the Activity Belief:}}
    It has been shown in~\cite{1971POMDP} that for each $t \in \mathcal{T}$ with $q_t= 1$, the value of the activity belief $\mathbf{b}_t$ is a sufficient statistic for the initial activity belief, all past channel statuses and all past transmission probabilities.
    First, by the total number of nodes $N$ and the packet generation probability $\lambda$, the tagged node can obtain
    \begin{align}
    \mathbf{b}_1 & = \mathbf{h}_\lambda \notag \\
    & \triangleq \big((1-\lambda)^{N-1}, (N-1)\lambda(1-\lambda)^{N-2},\ldots, \lambda^{N-1}\big). \label{initdist}
    \end{align}
    Then, for each $t \in \mathcal{T}\setminus\{D\}$, given the condition $q_t=q_{t+1}= 1$, the activity belief $\mathbf{b}_{t}=\mathbf{b}$, the observation $o_{t}=o$, the transmission probability $p_{t}=p$ used at slot $t$, the tagged node at slot $t+1$ can obtain $\mathbf{b}_{t+1}$ via the Bayes' rule:
    \begin{align}
    & \quad\quad\quad\quad\quad\,\,\,\,\,\,\,\,\,\, \mathbf{b}_{t+1}\triangleq \theta_t(\mathbf{b},p,o,1,1), \notag \\
    & b_{t+1}(n') \notag\\
    & \triangleq \text{Pr}\big(n_{t+1}=n'|\mathbf{b}_{t}=\mathbf{b},p_{t}=p,o_t=o,q_t=q_{t+1}= 1\big) \notag \\
    & = \frac{\sum_{n \in \mathcal{N}} b(n) \omega_t\big(o,(1,n),(1,n')\big) \beta_t\big((1,n'),(1,n),p\big)}{ \chi_t( o,\mathbf{b},p ,1,1) }, \notag
    \end{align}
    for each $n' \in \mathcal{N}$, where
    \begin{align}
        &\chi_t( o,\mathbf{b},p,1,1) \notag \\
        &\triangleq  \text{Pr}\big(q_{t+1}=1,o_t=o|q_t=1,\mathbf{b}_{t}=\mathbf{b},p_{t}=p\big) \notag\\
        &= \sum_{n \in \mathcal{N} } b(n) \sum_{n'' \in \mathcal{N}} \omega_t\big(o,(1,n),(1,n'')\big) \notag \\
        &\quad\quad\quad\quad\quad\quad\quad\,\, \cdot \beta_t\big((1,n''),(1,n),p\big). \notag
    \end{align}

    \item \underline{\emph{Policies:}}
    A deterministic Markovian policy $\bm{\pi}$ is defined by a sequence of transmission functions for the realistic environment:
     \begin{equation}
           \bm{\pi} \triangleq(\pi_1,\pi_2,\ldots,\pi_D),  \quad \text{where } \pi_t:  \mathcal{B}_t \rightarrow  [0,1]. \notag
    \end{equation}
    Let $\bm{\Pi}^{\text{MD}}$ denote the set of all possible such polices.
    Obviously, a dynamic control scheme for the realistic environment as specified in Section~\ref{sec:strategy} is essentially a deterministic Markovian policy here.
\end{enumerate}

Let $R^{\bm{\pi}}(1,\mathbf{h}_\lambda)$ denote the expected total reward from slot $1$ to slot $D$ when $q_1=1$, $\mathbf{b}_1 = \mathbf{h}_{\lambda}$ and the policy ${\bm{\pi}}$ is used, which can be defined by
\begin{align}
    & R^{\bm{\pi}}(1,\mathbf{h}_\lambda) \notag \\
    & \triangleq \mathbb{E}^{\bm{\pi}} \Big\{ \sum_{t=1, q_t=1}^D  r_t \big( (1,n_t),\pi_t(\mathbf{b}_t) \big) | q_1 =1, \mathbf{b}_1=\mathbf{h}_\lambda \Big\}. \notag
\end{align}
Obviously, we have $\text{TDR}^{\bm{\pi}}=R^{\bm{\pi}} (1,\mathbf{h}_\lambda)$, where $\text{TDR}^{\bm{\pi}}$ denotes the TDR under the policy $\bm{\pi}$,

\subsection{POMDP solution} \label{sec:pomdpsolution}
Due to the finite horizon, finite state space, compact action space, bounded rewards, continuous rewards with respect to $p$ and continuous $\chi_t(o,\mathbf{b},p,1,1)$ with respect to $p$ in our POMDP formulation,
\cite[Prop.~4.4.3, Ch.~4]{pm2014} and \cite[Thm.~7.1, Ch.~6]{kumar2015} indicate that for maximizing $\text{TDR}^{{\bm{\pi}}}$, there exists a ${\bm{\pi}} \in {\bm{\Pi}}^{\text{MD}}$, which is indeed optimal over all types of policies.
This property also justifies the transmission function and design goal for the realistic environment considered in Section~\ref{sec:strategy}.
Hence, we aim to seek an optimal policy in $\bm{\Pi}^{\text{MD}}$ that maximizes $\text{TDR}^{\bm{\pi}}$, i.e.,
\begin{equation}
   \bm{\pi}^* \in \mathop{\arg\max}_{\bm{\pi} \in \bm{\Pi}^{\text{MD}}} \text{TDR}^{\bm{\pi}}. \notag
\end{equation}

Let $V^*_t(1,\mathbf{b})$ denote the value function corresponding to the maximum total expected reward from slot $t$ to slot $D$ when $q_t=1$ and $\mathbf{b}_t = \mathbf{b}$. Averaging over all possible current states with $q_{t}=1$ and observations with $q_{t+1}=1$,
we arrive at the following Bellman's equation:
\begin{equation}\label{BellmanII}
\begin{aligned}
     & V^*_D(1,\mathbf{b}) = \mathop{\max}_{p \in [0,1]}   \sum_{n\in \mathcal{N} } b(n) r_D\big((1,n),p\big), \quad \forall \mathbf{b} \in \mathcal{B}_D,\\
     & V^*_t(1,\mathbf{b}) = \mathop{\max}_{p \in [0,1]}   \sum_{n\in \mathcal{N} } b(n) r_t\big((1,n),p\big) \\
     & \quad + \sum_{o \in \mathcal{O}} \chi_t( o,\mathbf{b},p,1,1) V^*_{t+1}\big(1,\theta_t(\mathbf{b},p,o,1,1)\big), \quad  \forall \mathbf{b} \in \mathcal{B}_t,
\end{aligned}
\end{equation}
for each $t\in \mathcal{T}\setminus \{D\}$.
Solving Eq.~\eqref{BellmanII} formally leads to $\bm{\pi}^*$.

Unfortunately, getting $\bm{\pi}^*$ by solving Eq.~\eqref{BellmanII} is computationally intractable, as both the belief state space $\bigcup_{t\in \mathcal{T}}\mathcal{B}_t$ and the action space $[0,1]$ are infinite in our POMDP formulation.
As such, an alternative is to consider a discretized action space $\mathcal{A}_d$ that only consists of uniformly distributed samples of the interval $[0,1]$, i.e., $\mathcal{A}_d \triangleq\{ 0,\Delta p,2\Delta p,\ldots,1 \}$ where $\Delta p$ denotes the sampling interval.
Hence, it is easy to see that $\mathcal{B}_t$ will become finite for each $t\in \mathcal{T}$ due to the finite $\mathcal{A}_d$.
Then, theoretically, applying the backward induction algorithm~\cite{1971POMDP} to get a solution to Eq.~\eqref{BellmanII} can lead to a near-optimal policy, whose loss of optimality increases with $\Delta p$.
However, this approach is still computationally prohibitive due to super-exponential
growth in the value function complexity.

\section{A Heuristic Scheme for the Realistic Environment}  \label{sec:heuristicstrategy}

To overcome the infeasibility in obtaining an optimal or near-optimal control scheme for the realistic environment from the POMDP framework, in this section, we propose a simple heuristic control scheme that utilizes the key properties of our problem.
It will be shown in Section~\ref{sec:numerical} that the heuristic scheme performs quite well in simulations.

\subsection{Heuristic from the idealized environment} \label{sec:heuridealized}

We first investigate the behaviors of $\widehat{\bm{\pi}}^*$ for two extreme cases in the idealized environment, which would serve to provide important clues on approximating $\widehat{\bm{\pi}}^*$.
Let $U_t\big((1,n),p\big)$ denote the total expected reward from slot $t$ to slot $D$ for the state $(q_t,n_t) = (1,n)$ when each active node at the beginning of slot $t$ adopts the transmission probability $p_t = p$ and the optimal decision rules at slots $t+1,t+2,\ldots,D$.
So, we have $U_t\big((1,n),\widehat{\pi}^*_t(n)\big)=U^*_t (1,n)$.

\begin{lemma}
When $n_1=m\rightarrow\infty$, by assuming that each collision involves a finite number of packets, for each $t \in \mathcal{T}$ and each possible $n_t=n$, we have
\begin{align}
  & \lim_{ m \to \infty} (n+1)  U_t \big((1,n),\frac{1}{n+1}\big) = \dfrac{(D-t+1)\sigma}{e}, \label{lemma1} \\
  & \lim_{ m \to \infty} (n+1)  U^*_t (1,n) = \dfrac{(D-t+1)\sigma}{e}. \label{lemma12}
\end{align}
\end{lemma}

The proof of Lemma 1 is provided in Appendix A.

\begin{figure}[!ht]
	\centering
	\includegraphics[width=3.4in]{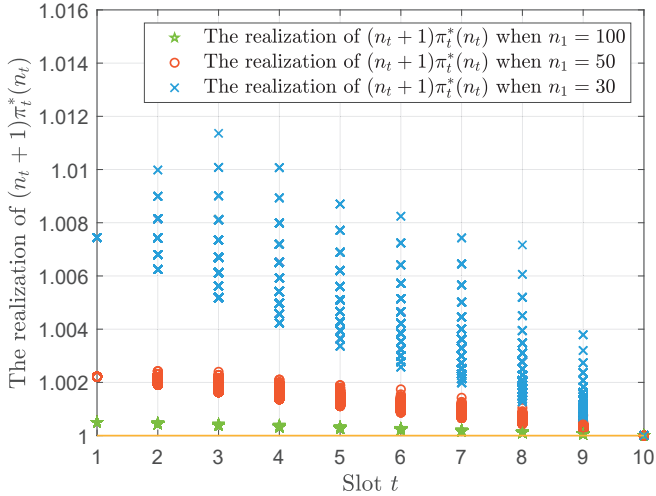}
	\caption{Realizations of $(n_t+1)\widehat{\pi}^*_t (n_t)$ when $n_1=30, 50, 100$ and $D=10$.}
    \label{realization-1}
\end{figure}

Lemma 1 motivates us to conjecture that, if $n_1$ takes a value sufficiently larger than $D$, the realizations of $(n_t+1)\widehat{\pi}^*_t (n_t)$ would always approach 1 for each $t \in \mathcal{T}$.
Fig.~\ref{realization-1} shows 1000 such realizations when $D=10$ for $n_1=30, 50, 100$, respectively, which confirm our conjecture.

We further investigate the behaviors of $\widehat{\bm{\pi}}^*$ for the extreme case that $n_1$ takes a value sufficiently smaller than $D$.

\begin{lemma}
For each $t \in \mathcal{T}$,  we have
\begin{equation}\label{lemma21}
    U^*_t(1,1) = \frac{3D-3t+1}{3D-3t+4}\sigma,
\end{equation}
and for each $t \in \mathcal{T} \setminus \{D\}$, we have
\begin{equation}\label{lemma22}
    \widehat{\pi}^*_t(1) = \frac{3}{ 3D-3t+4 }.
\end{equation}
\end{lemma}

The proof of Lemma 2 is provided in Appendix B.

Inspired by Eq.~\eqref{lemma22}, we consider a simple control scheme $\widehat{\bm{\pi}}^\text{eve} \triangleq [\widehat{\pi}^\text{eve}_1,\widehat{\pi}^\text{eve}_2,\ldots,\widehat{\pi}^\text{eve}_D] \in \widehat{\bm{\Pi}}^{\text{MD}}$ where
\begin{align}
    \widehat{\pi}^\text{eve}_t(n) = \frac{1}{D-t+1},
\end{align}
for each $t \in \mathcal{T}$ and each $n \in \mathcal{N}$.

Let $U^{\text{eve}}_t(1,n)$ denote the expected total reward from slot $t$ to slot $D$ for the state $(q_t,n_t) = (1,n)$ when each active node adopts the decision rules $\widehat{\pi}^\text{eve}_t$ at slots $t,t+1,\ldots,D$.
So, using the finite-horizon policy evaluation algorithm~\cite{pm2014}, we have
\begin{equation} \label{eve-bellman}
\begin{aligned}
     & U^{\text{eve}}_D(1,n) = r_D \big((1,n),\widehat{\pi}_D^{\text{eve}}(n)\big),  \quad \forall n \in \mathcal{N},\\
     & U^{\text{eve}}_t(1,n) = r_t \big((1,n),\widehat{\pi}_t^{\text{eve}}(n)\big) \\
     & \quad + \sum_{n' \in \mathcal{N}} \beta_t \big((1,n'),(1,n),\widehat{\pi}_t^{\text{eve}}(n)\big) U^{\text{eve}}_{t+1}(1,n'), \ \forall n \in \mathcal{N}.
\end{aligned}
\end{equation}
for each $t \in \mathcal{T}\setminus \{D\}$.

\begin{lemma}
For each $t \in \mathcal{T} \setminus \{D\}$ and each $n \in \mathcal{N}$,  we have
\begin{align}
    U^{\text{eve}}_t(1,n) = \sigma \big(1-\frac{1}{D-t+1}\big)^n. \label{even-value}
\end{align}
\end{lemma}

The proof of Lemma 3 is provided in Appendix C.

\begin{figure}[!ht]
	\centering
	\includegraphics[width=3.4in]{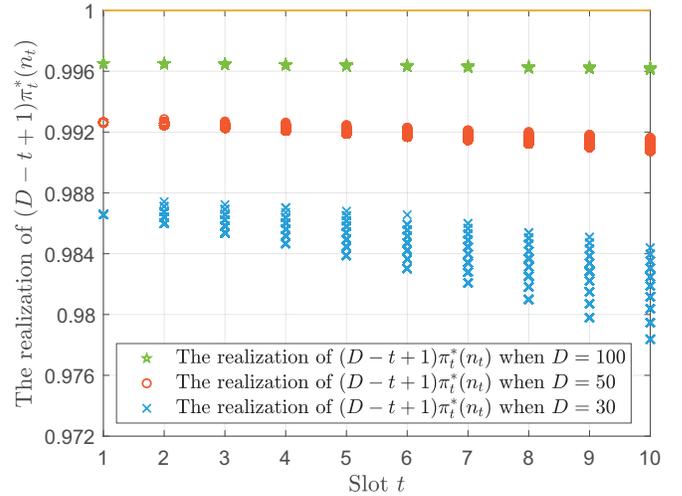}
	\caption{Realizations of $(D-t+1)\widehat{\pi}^*_t (n_t)$ when $n_1=10$ and $D=30,50,100$.}
    \label{realization-2}
\end{figure}

For each $t \in \mathcal{T} \setminus \{D\}$ and each $n \in \mathcal{N}$, based on the fact $U^{*}_t(1,n) \leq \sigma$,  we have
\begin{align}
   \frac {U^{\text{eve}}_t(1,n)} {U^*_t(1,n)}  \geq  \big(1-\frac{1}{D-t+1}\big)^n. \label{opt-even}
\end{align}
We can observe from Eq.~\eqref{opt-even} that, if $n$ is sufficiently smaller than $D-t+1$, the value of $U^*_t(1,n)$ is close to the value of $U^{\text{eve}}_t(1,n)$.
Then, Eq.~\eqref{opt-even} motivates us to conjecture that, if $n_t$ takes a value sufficiently smaller than $D-t+1$, $\widehat{\pi}^\text{eve}_t$ may behave like $\widehat{\pi}^*_t$, i.e., the realizations of $(D-t+1)\widehat{\pi}^*_t (n_t)$ would always approach $(D-t+1)\widehat{\pi}^\text{eve}_t (n_t) = 1$ for each $n_t = n \in \mathcal{N}$.
Fig.~\ref{realization-2} shows 1000 such realizations when $n_1 = 10$ for $D = 30,50,100$, respectively, which confirm our conjecture.

Naturally, we obtain the following heuristic from Lemma 1 and Eq.~\eqref{opt-even}.
\begin{enumerate}
  \item When the number of active nodes is sufficiently large compared with the value of remaining slots, it is desirable for the active nodes to adopt the transmission probability that maximizes the instantaneous throughput.

  \item When the number of active nodes is sufficiently small compared with the value of remaining slots, it is desirable for the active nodes to adopt the transmission probability to ensure that all the backlogged packets would be almost evenly transmitted in the remaining slots.
\end{enumerate}
Based on this heuristic and the obvious fact $\widehat{\pi}^*_D(n) = \frac{1}{n+ 1}$ for each $n\in \mathcal{N}$, we propose a simple approximation on $\widehat{\bm{\pi}}^*$.

\medskip
\noindent\emph{Approximation on $\widehat{\bm{\pi}}^*$: For each slot $t \in \mathcal{T}$ and each $n_t=n\in \mathcal{N}$, if the number of active nodes $n+1$ is larger than the value of remaining slots $D-t+1$ or $t = D$, $\widehat{\pi}^*_t(n)$ can be estimated by $\frac{1}{n+1}$, otherwise $\widehat{\pi}^*_t(n)$ can be estimated by $\frac{1}{ D - t + 1 }$.}
\medskip

\begin{figure}[!ht]
	\centering
	\includegraphics[width=3.4in]{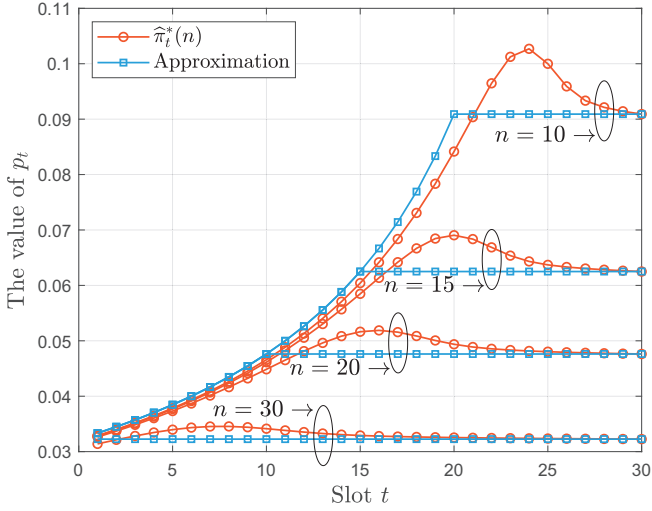}
	\caption{$\widehat{\pi}^*_t(n)$ and its approximation for typical choices of parameters when $D=30$.}
    \label{heurp}
\end{figure}

\begin{figure}[!ht]
	\centering
	\includegraphics[width=3.4in]{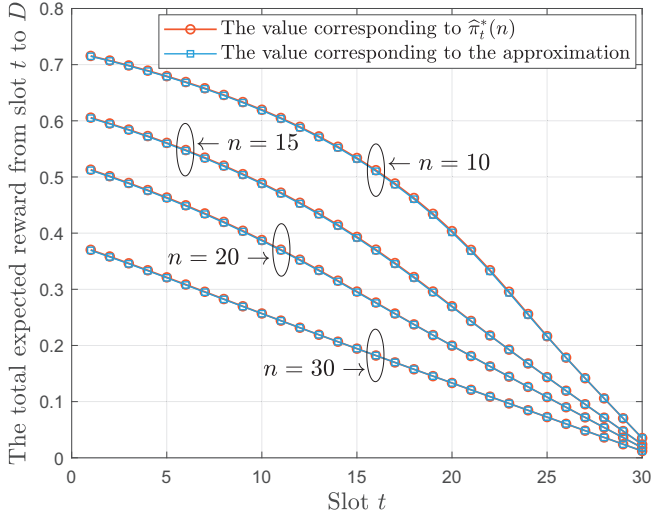}
	\caption{The total expected rewards from slot $t$ to $D$ corresponding to $\widehat{\pi}^*_t(n)$ and its approximation for typical choices of parameters when $D=30$.}
    \label{heuru}
\end{figure}

Fig.~\ref{heurp} compares $\widehat{\pi}^*_t(n)$ and its approximation for typical choices of parameters when $D=30$.
The results show that the approximation error is very small when the difference between $n$ and $D-t$ is large, but is noticeable when this difference is small, thus justifying our heuristic.
The results also show that the ratio of the cases with the approximation error larger than $8\%$ is $6.67\%$ and the largest approximation error is $11.47\%$, thus justifying our approximation.
Furthermore, Fig.~\ref{heuru} compares the total expected rewards from slot $t$ to $D$ corresponding to $\widehat{\pi}^*_t(n)$ and its approximation for typical choices of parameters when $D=30$.
The results show that the approximation leads to at most $0.66\%$ reward loss (much smaller than the approximation error), and thus verify that our approximation can be used to obtain the TDR quite close to the maximum achievable TDR in the idealized environment.

\begin{table*}
\caption{A comparison between the realizations of $\mathbf{b}_t$ and its approximation ${\mathbf{b}}_t^{\text{\upshape bd}}$ when each active node adopts $\bm{\pi}^{\text{\upshape heu}}$ for $N = 10$, $\lambda = 0.8$, $D = 10$.}
\begin{tabular}{|c|c|c|c|c|c|c|c|c|c|c|c|} \hline
                       &   & $b_t(0)$  & $b_t(1)$  & $b_t(2)$  & $b_t(3)$  & $b_t(4)$  & $b_t(5)$  & $b_t(6)$  & $b_t(7)$  & $b_t(8)$  & $b_t(9)$  \\ \hline\hline
\multicolumn{1}{|c|}{\multirow{2}{*}{\begin{tabular}[c]{@{}c@{}} $t = 1$ \\ $o_1 = 0$\end{tabular}}} & $\mathbf{b}_t$ & 0.000001 & 0.000018 & 0.000295 & 0.002753 & 0.016515 & 0.066060 & 0.176161 & 0.301990 & 0.301990 & 0.134218        \\ \cline{2-12}
                       & Approx. & 0.000001 & 0.000018 & 0.000295 & 0.002753 & 0.016515 & 0.066060 & 0.176161 & 0.301990 & 0.301990 & 0.134218 \\ \hline\hline
\multicolumn{1}{|c|}{\multirow{2}{*}{\begin{tabular}[c]{@{}c@{}} $t = 2$ \\ $o_2 = 1$\end{tabular}}} & $\mathbf{b}_t$ & 0.000001 & 0.000042 & 0.000583 & 0.004760 & 0.024988 & 0.087458 & 0.204068 & 0.306102 & 0.267839 & 0.104160        \\ \cline{2-12}
                       & Approx. & 0.000001 & 0.000042 & 0.000583 & 0.004760 & 0.024988 & 0.087458 & 0.204068 & 0.306102 & 0.267839 & 0.104160        \\ \hline\hline
\multicolumn{1}{|c|}{\multirow{2}{*}{\begin{tabular}[c]{@{}c@{}} $t = 3$ \\ $o_3 = 1$\end{tabular}}} & $\mathbf{b}_t$ & 0.000059 & 0.001098 & 0.009014 & 0.042646 & 0.127254 & 0.245406 & 0.298859 & 0.210235 & 0.065430 & 0        \\ \cline{2-12}
                       & Approx. & 0.000052 & 0.001004 & 0.008559 & 0.041692 & 0.126924 & 0.247294 & 0.301138 & 0.209545 & 0.063792 & 0        \\ \hline\hline
\multicolumn{1}{|c|}{\multirow{2}{*}{\begin{tabular}[c]{@{}c@{}} $t = 4$ \\ $o_4 = 1$\end{tabular}}} & $\mathbf{b}_t$ & 0.001086 & 0.012248 & 0.059916 & 0.164987 & 0.276437 & 0.282086 & 0.162465 & 0.040774 & 0        & 0        \\ \cline{2-12}
                       & Approx. & 0.000974 & 0.011537 & 0.058598 & 0.165343 & 0.279925 & 0.284347 & 0.160466 & 0.038810 & 0        & 0        \\ \hline\hline
\multicolumn{1}{|c|}{\multirow{2}{*}{\begin{tabular}[c]{@{}c@{}} $t = 5$ \\ $o_5 = 1$\end{tabular}}} & $\mathbf{b}_t$ & 0.010921 & 0.072058 & 0.201100 & 0.304173 & 0.263268 & 0.123764 & 0.024716 & 0        & 0        & 0        \\ \cline{2-12}
                       & Approx. & 0.010329 & 0.070827 & 0.202359 & 0.308353 & 0.264299 & 0.120821 & 0.023013 & 0        & 0        & 0        \\ \hline\hline
\multicolumn{1}{|c|}{\multirow{2}{*}{\begin{tabular}[c]{@{}c@{}} $t = 6$ \\ $o_6 = 0$\end{tabular}}} & $\mathbf{b}_t$ & 0.068102 & 0.238724 & 0.340491 & 0.247285 & 0.091556 & 0.013842 & 0        & 0        & 0        & 0        \\ \cline{2-12}
                       & Approx. & 0.067210 & 0.240606 & 0.344541 & 0.246686 & 0.088312 & 0.012646 & 0        & 0        & 0        & 0        \\ \hline\hline
\multicolumn{1}{|c|}{\multirow{2}{*}{\begin{tabular}[c]{@{}c@{}} $t = 7$ \\ $o_7 = 0$\end{tabular}}} & $\mathbf{b}_t$ & 0.169904 & 0.357679 & 0.306377 & 0.133629 & 0.029713 & 0.002698 & 0        & 0        & 0        & 0        \\ \cline{2-12}
                       & Approx. & 0.167239 & 0.359554 & 0.309208 & 0.132956 & 0.028585 & 0.002458 & 0        & 0        & 0        & 0        \\ \hline\hline
\multicolumn{1}{|c|}{\multirow{2}{*}{\begin{tabular}[c]{@{}c@{}} $t = 8$ \\ $o_8 = 1$\end{tabular}}} & $\mathbf{b}_t$ & 0.421334 & 0.395352 & 0.150943 & 0.029344 & 0.002908 & 0.000118 & 0        & 0        & 0        & 0        \\ \cline{2-12}
                       & Approx. & 0.416144 & 0.398784 & 0.152859 & 0.029297 & 0.002807 & 0.000108 & 0        & 0        & 0        & 0        \\ \hline
\end{tabular}
\label{table:beliefapprox}
\end{table*}
\subsection{A simple approximation on the activity belief of the realistic environment} \label{sec:estimation}

To apply the approximation on $\widehat{\bm{\pi}}^*$ to the realistic environment, it is necessary for each active node to perform a runtime updating of the activity belief $\mathbf{b}_t$.
However, as shown in Section \ref{sec:pomdpformulation}, the full Bayesian updating of $\mathbf{b}_t$ is a bit computationally demanding to implement.
So, we shall propose a simple approximation on $\mathbf{b}_t$, denoted by ${\mathbf{b}}_t^{\text{bd}} \triangleq \big({b}_t^{\text{bd}}(0), {b}_t^{\text{bd}}(1),\ldots, {b}^{\text{bd}}_t(N-1)\big)$, relying on a binomial distribution with a changeable parameter vector $(M_t,\alpha_t)$.
More specifically, if $(M_t,\alpha_t)=(M,\alpha)$, we have
\begin{align}
    {b}_t^{\text{bd}}(n)=
    \begin{cases}
    \binom{M}{n}\alpha^{n}(1-\alpha)^{M-n},  &\text{if}\,\, 0 \leq n \leq M, \\
    0,  &\text{otherwise}. \label{beliefbinom}
    \end{cases}
\end{align}
As such, in this manner, each active node will only keep the parameter vector $(M_t,\alpha_t)$ rather than the activity belief $\mathbf{b}_t$.

Obviously, by Eq.~\eqref{initdist}, we can set $(M_1,\alpha_1)=(N-1,\lambda)$ to achieve ${\mathbf{b}}_t={\mathbf{b}}_t^{\text{bd}}$.
Then, for each $t \in \mathcal{T}\setminus\{D\}$, we will show that we can use the Bayes' rule exactly to set the value of $(M_{t+1},\alpha_{t+1})$ when the observation $o_t = 0$, but must introduce an approximation assumption when $o_t = 1$.

For each $t \in \mathcal{T}\setminus\{D\}$, given $(M_t,\alpha_t)=(M,\alpha)$, $\mathbf{b}_t^{\text{bd}} = {\mathbf{b}}^{\text{bd}}$ and $p_t=p$, the following procedure first uses the Bayes' rule to compute ${\mathbf{b}}_{t+1}^{\text{med}} \triangleq \big({b}_{t+1}^{\text{med}}(0), {b}_{t+1}^{\text{med}}(1),\ldots, {b}_{t+1}^{\text{med}}(N-1)\big)$, and then computes $(M_{t+1},\alpha_{t+1})$ based on the value of ${\mathbf{b}}_{t+1}^{\text{med}}$.

\noindent \emph{Case 1:} if $o_t = 0$, the Bayesian update yields
\begin{align}
    & b_{t+1}^{\text{med}}(n') \notag \\
    & = \frac{\sum_{n \in \mathcal{N}} {b}^{\text{bd}}(n) \omega_t\big(0,(1,n),(1,n')\big) \beta_t\big((1,n'),(1,n),p\big)}{\chi_t( 0,{\mathbf{b}}^{\text{bd}},p ,1,1) } \notag \\
    & =
    \begin{cases}
        \binom{M}{n'} \Big(\frac{\alpha - \alpha p}{1 - \alpha p}\Big)^{n'}\Big( 1 - \frac{\alpha - \alpha p}{1 - \alpha p}\Big)^{M-n'}, &\text{if}\,\, 0 \leq n' \leq M, \\
        0,  &\text{otherwise}.
    \end{cases} \notag
\end{align}
We require $\mathbf{b}^{\text{bd}}_{t+1}$ to directly take the value of ${\mathbf{b}}_{t+1}^{\text{med}}$, and set
\[
(M_{t+1},\alpha_{t+1}) = (M, \frac{\alpha - \alpha p}{1 - \alpha p}).
\]

\noindent \emph{Case 2:} if $o_t = 1$, the Bayesian update yields
    \begin{align}
        & b_{t+1}^{\text{med}}(n') \notag \\
        & = \frac{\sum_{n \in \mathcal{N}} b^{\text{bd}}(n) \omega_t\big(1,(1,n),(1,n')\big) \beta_t\big((1,n'),(1,n),p\big)}{\chi_t( 1, \mathbf{b}^{\text{bd}} ,p ,1,1)} \notag \\
        & =\begin{cases}
        \frac{1}{1 - (1 - \alpha p)^{M}}& \notag \\
        \cdot \bigg[ \binom{M}{n'} \big(\alpha(1-p)\big)^{n'} \big( 1 - \alpha(1-p) \big)^{M-n'} \notag \\
        \quad- (1-\alpha p)^M  \binom{M}{n'} \Big(\frac{\alpha - \alpha p}{1 - \alpha p}\Big)^{n'}\Big( 1 - \frac{\alpha - \alpha p}{1 - \alpha p}\Big)^{M-n'}\bigg], \notag \\
        \quad\quad\quad\quad\quad\quad\quad\quad \quad \quad\quad\quad\quad\quad \text{if}\,\, 0 \leq n' \leq M-1,  & \notag \\
        0,\quad\quad\quad\quad\quad\quad\quad\quad\quad\quad\quad\quad\quad \text{otherwise}. \notag
    \end{cases}
    \end{align}
     Such a Bayesian update does not yield a distribution in the form~\eqref{beliefbinom}.
    However, we modify the value of ${\mathbf{b}}_{t+1}^{\text{med}}$ to a distribution in the form~\eqref{beliefbinom} by keeping the mean of the distribution unchanged and considering that the number of active nodes will be reduced by at least one due to a busy slot.
    So, when $M>1$, we set
    \begin{align}
        &(M_{t+1},\alpha_{t+1})= \bigg(M - 1, \frac{M(\alpha - \alpha p)\big(1 - (1 - \alpha p)^{M-1} \big)} {(M-1)  \big(1 - (1 - \alpha p)^{M} \big)} \bigg), \notag
    \end{align}
    and when $M=1$, we adopt the convention that
    \[
    (M_{t+1},\alpha_{t+1})= (M - 1, 1).
    \]

The accuracy of this approximation will be examined via numerical results at the end of this section.

\subsection{A heuristic scheme} \label{sec:heurrealistic}
With the investigations in Sections~\ref{sec:heuridealized} and~\ref{sec:estimation} together, we are ready to propose a heuristic but very simple control scheme for the realistic environment, ${\bm{\pi}}^{\text{heu}}$.

At the beginning of each slot $t \in \mathcal{T}$, given the parameter vector of belief approximation $(M_t,\alpha_t) = (M,\alpha)$, each active node uses $M\alpha$ to estimate the mean of $n_t$, and further uses the following simple rule ${\pi}^{\text{heu}}_t(M,\alpha)$ to determine the value of transmission probability $p_t$.
\begin{enumerate}
  \item If $M\alpha + 1 > D-t + 1$ or $t = D$, set ${\pi}^{\text{heu}}_t(M,\alpha)$ to maximize the expected instantaneous throughput, i.e.,
  \begin{align}
  {\pi}^{\text{heu}}_t( M,\alpha ) &\in \mathop{\arg\max}_{p \in [0,1]} \sum_{n \in \mathcal{N}} {b}^{\text{bd}}(n) r_t\big((1,n),p\big). \notag \\
  & = \min \bigg( \frac{1}{M\alpha + \alpha},1 \bigg). \label{heur_immd}
  \end{align}
  The proof of Eq.~\eqref{heur_immd} is provided in Appendix D.
  \item Otherwise, set
    \begin{equation}
         {\pi}^{\text{heu}}_t(M,\alpha) =\frac{ 1 } { D - t + 1 }. \notag
    \end{equation}
\end{enumerate}

Table~\ref{table:beliefapprox} compares the realizations of $\mathbf{b}_t$ and its approximation ${\mathbf{b}}_t^{\text{bd}}$ when each active node adopts $\bm{\pi}^{\text{heu}}$ for $N = 10$, $\lambda = 0.8$, $D = 10$, and verifies that the proposed approximation is reasonable.

% [1;1;3;2;3;1;1;2;1;2]

\section{Numerical Evaluation} \label{sec:numerical}

%This section includes two subsections.
In this section, we present numerical results to compare the TDR performance of an optimal control scheme for the idealized environment $\widehat{\bm{\pi}}^*$, the proposed heuristic scheme for the realistic environment ${\bm{\pi}}^{\text{heu}}$, and an optimal static scheme for the realistic environment ${\bm{\pi}}^{\text{sta}}$.
Here, ${\bm{\pi}}^{\text{sta}}$ requires each active node to always adopt a static and identical transmission probability, and can be obtained using the single-variable optimization methods.
Such compactions are not only helpful to demonstrate the performance loss due to the incomplete knowledge of the value of $n_t$, but also helpful to demonstrate the performance advantage benefitting from dynamically adjusting the transmission probability.

\begin{figure}[!ht]
	\centering
	\includegraphics[width=3.4in]{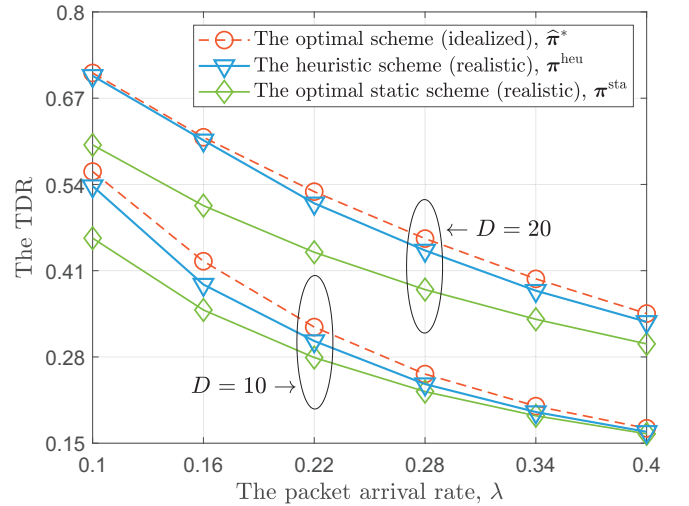}
	\caption{The TDR as a function of the packet arrival rate $\lambda$ for $N = 50$, $D=10, 20$, $\sigma = 0.9$.}
    \label{fig-lambda}
\end{figure}

The scenarios considered in the numerical experiments are in accordance with the system model specified in Section~\ref{sec:systemstrategy}.
%The packet success rate due to channel errors will be specified for each scenario.
We shall vary the network configuration over a wide range to investigate the impact of control scheme design on the TDR performance.
Each numerical result is obtained from $10^7$ independent numerical experiments.

Fig.~\ref{fig-lambda} shows the TDR performance as a function of the packet arrival rate $\lambda$ for $N=50$, $D=10, 20$, $\sigma = 0.9$.
We observe that ${\bm{\pi}}^{\text{heu}}$ performs close to $\widehat{\bm{\pi}}^*$: $3.07\%$--$8.28\%$ loss when $D=10$ and $0.60\%$--$4.47\%$ loss when $D=20$.
This indicates that the design of ${\bm{\pi}}^{\text{heu}}$ is reasonable, and the incomplete knowledge of the actual value of $n_t$ has a minor impact on the TDR performance.
We further observe that ${\bm{\pi}}^{\text{heu}}$ significantly outperforms ${\bm{\pi}}^{\text{sta}}$: $1.84\%$--$17.12\%$ improvement when $D=10$ and $11.11\%$--$19.40\%$ improvement when $D=20$.
The reason is obviously that ${\bm{\pi}}^{\text{sta}}$ does not adjust the transmission probability according to the current delivery urgency and contention intensity.
Meanwhile, it is interesting to note that ${\bm{\pi}}^{\text{sta}}$ performs closer to other schemes as $\lambda$ increases.
This is because the optimal transmission probabilities for different values of $t$ and $n_t$ become closer with the value of $n_1$, as indicated by Lemma 1.

\begin{figure}[!ht]
	\centering
	\includegraphics[width=3.4in]{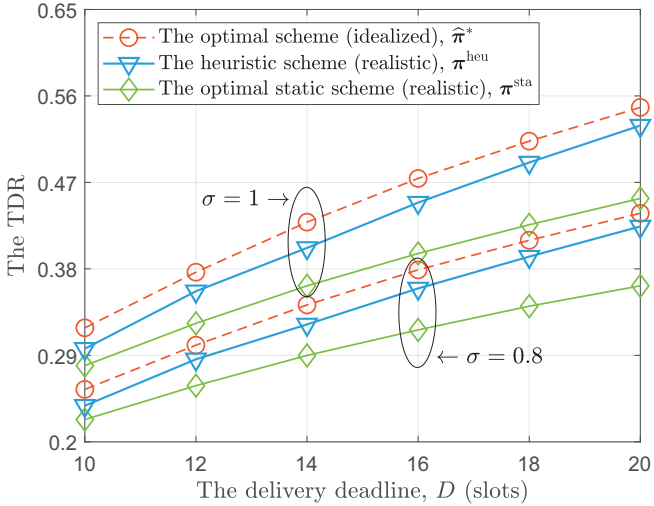}
	\caption{The TDR as a function of the delivery deadline $D$ (slots) for $N = 50$, $\lambda=0.25$, $\sigma = 0.8,1$.}
    \label{fig-D}
\end{figure}

\begin{figure}[!ht]
	\centering
	\includegraphics[width=3.4in]{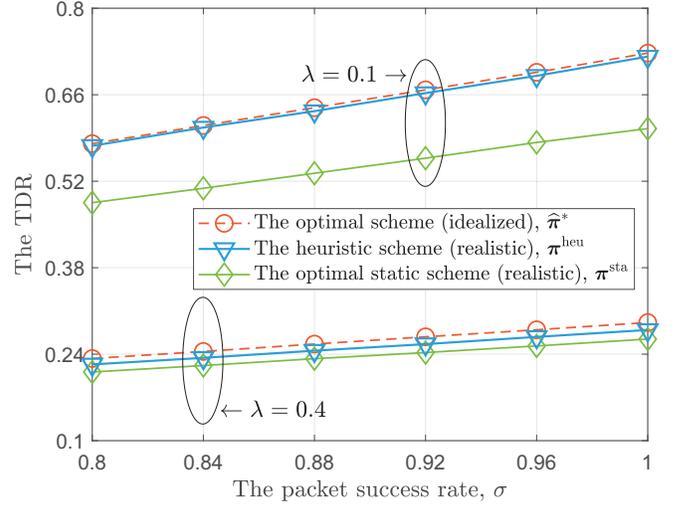}
	\caption{The TDR as a function of the packet success rate $\sigma$ for $N = 50$, $\lambda=0.1,0.4$, $D = 15$.}
    \label{fig-sigma}
\end{figure}

The observations in Fig.~\ref{fig-lambda} are confirmed again from Figs.~\ref{fig-D}--\ref{fig-sigma}, which show the TDR performance as a function of the delivery deadline $D$ (slots) and the TDR performance as a function of the packet success rate $\sigma$, respectively.
In Fig.~\ref{fig-D}, we observe that ${\bm{\pi}}^{\text{heu}}$ performs significantly better than ${\bm{\pi}}^{\text{sta}}$: $6.45\%$--$17.06\%$ improvement when $\sigma=0.8$ and $6.30\%$--$16.74\%$ improvement when $\sigma=1$, and performs close to $\widehat{\bm{\pi}}^*$: $3.12\%$--$6.68\%$ loss when $\sigma=0.8$ and $3.45\%$--$6.83\%$ loss when $\sigma=1$.
In Fig.~\ref{fig-sigma}, we observe that ${\bm{\pi}}^{\text{heu}}$ performs significantly better than ${\bm{\pi}}^{\text{sta}}$: $18.51\%$--$19.33\%$ improvement when $\lambda = 0.1$ and $5.58\%$--$5.81\%$ improvement when $\lambda = 0.4$, and performs close to $\widehat{\bm{\pi}}^*$: $0.55\%$--$0.87\%$ loss when $\lambda = 0.1$ and $4.11\%$--$4.41\%$ loss when $\lambda = 0.4$.
It is also shown that ${\bm{\pi}}^{\text{sta}}$ performs close to other transmission schemes as $\frac{N\lambda}{D}$ becomes larger.

\section{Conclusion} \label{Conclusion}

In this paper, under the idealized and realistic environments, optimal dynamic control schemes for random access in deadline-constrained broadcasting with frame-synchronized traffic have been investigated based on the theories of MDP and POMDP, respectively.
A novel feature of this work is to require each active node to determine the current transmission probability not only according to the knowledge of current contention intensity, but also according to the current delivery urgency.
The proposed heuristic scheme for the realistic environment is able to achieve the threefold goal of being implemented without imposing extra overhead and hardware cost, of being implemented with very low computational complexity, and of achieving TDR close to the maximum achievable TDR in the idealized environment.
%Another heuristic scheme with similar features is proposed for the realistic environment with channel errors.
%Furthermore, it has been shown that our study can be easily extended to incorporate random fading effect.
%consider multi-slot packets and other stochastic processes of packet arrivals under frame-synchronized traffic.
An interesting and important future research direction is to optimize deadline-constrained broadcasting under general traffic patterns.

\appendices
\section{Proof of Lemma 1}
Assume each collision involves at most a finite number, $k\geq 2$, of packets.
We begin with the case ${t = D}$.
By Eq.~\eqref{reward}, we know $U_D\big((1,n),p\big) = \sigma p(1-p)^{n}$ and thus $\widehat{\pi}^*_D (n) = \frac{1}{n+1}$.
As $k$ and $D$ are both finite, for each $n_D=n \in \{m-k(D-1),m-k(D-1)+1,\ldots, m\}$, we obtain that $m \to \infty$ implies $n \to \infty$ and then
\begin{equation} \label{t-D}
    \lim_{ m \to \infty} (n+1)U^*_D(1,n) = \lim_{ m \to \infty} (n+1)U_D \big((1,n),\frac{1}{n+1}\big) = \frac{\sigma}{e}.
\end{equation}

Next, we consider the case ${t = D - 1}$.
By the finite-horizon policy evaluation algorithm~\cite{pm2014} and Eqs.~\eqref{trans}, \eqref{reward}, for each $n_{D-1} = n\in \{m-k(D-2),m-k(D-2)+1, \ldots, m\}$, we have
\begin{align}
    &(n+1)U_{D-1}\big((1,n),p\big) \notag \\
    & = (n+1)r_{D-1}\big((1,n),p\big) \notag \\
    & \quad + (n+1) \sum_{n' \in \mathcal{N}} \beta_{D-1}\big((1,n'),(1,n),p\big) U^*_{D}(1,n') \notag\\
    & = \sigma (n+1)p(1-p)^n \notag \\
    & \quad + \sum_{n' \in \mathcal{N}} (n+1) \frac{n!}{n'!(n-n')!} p^{n-n'}(1-p)^{n'+1} U^*_{D}(1,n') \notag \\
    & = \sigma (n+1) p(1-p)^n \notag \\
    & \quad + \sum_{n' \in \mathcal{N}} \binom{n+1}{n-n'} p^{n-n'}(1-p)^{n'+1} (n'+1)U^*_{D}(1,n'). \notag
\end{align}
By assuming each collision involves at most a finite number, $k\geq 2$, of packets, we have
\begin{align}
    & (n+1)U_{D-1}\big((1,n),p\big) \notag \\
    & = \sigma (n+1) p(1-p)^n \notag \\
    & + \sum_{n' = n-k+1}^{n} \binom{n+1}{n-n'} p^{n-n'}(1-p)^{n'+1} (n'+1)U^*_{D}(1,n') \notag \\
    & + \Big(1 - \sum_{n' = n-k+1}^{n} \binom{n+1}{n-n'} p^{n-n'}(1-p)^{n'+1} \Big) \notag \\
    & \quad \cdot (n-k+1)U^*_{D}(1,n-k) \label{approx_1} \\
    & \leq \sigma (1-\frac{1}{n+1})^n + (n-k+1)U^*_{D}(1,n-k) \notag \\
    & + \sum_{n' = n-k+1}^{n} \binom{n+1}{n-n'} p^{n-n'}(1-p)^{n'+1} \notag \\
    & \quad \cdot \big( (n'+1)U^*_{D}(1,n') - (n-k+1)U^*_{D}(1,n-k) \big). \label{approx_2}
\end{align}
For each $n' \in \{n-k+1,n-k+2,\ldots,n\}$, since $0 \leq \binom{n+1}{n-n'} p^{n-n'}(1-p)^{n'+1} \leq 1$, by applying the squeeze theorem, we obtain from Eq.~\eqref{t-D} that
\begin{align} \label{squeeze}
     & \lim_{m \to \infty} \binom{n+1}{n-n'} p^{n-n'}(1-p)^{n'+1} \notag \\
     & \quad\quad \cdot \big( (n'+1)U^*_{D}(1,n') - (n-k+1)U^*_{D}(1,n-k) \big) = 0.
\end{align}
By Eqs.~\eqref{t-D}, \eqref{squeeze} and inequality~\eqref{approx_2}, as $k$ and $D$ are both finite, we further obtain that $m \to \infty$ implies $n \to \infty$ and then
\begin{align}
     &\mathop{\lim\sup}_{m \to \infty} (n+1)U_{D-1}\big((1,n),p\big) \notag \\
    &  \leq \mathop{\lim\sup}_{m \to \infty} \big( \sigma (1-\frac{1}{n+1})^n + (n-k+1)U^*_{D}(1,n-k) \big) \notag \\
      &  = \lim_{m \to \infty} \big( \sigma (1-\frac{1}{n+1})^n + (n-k+1)U^*_{D}(1,n-k) \big) \notag \\
      &= \frac{2\sigma}{e}, \notag
\end{align}
which implies
\begin{equation} \label{limsup}
    \mathop{\lim\sup}_{m \to \infty} (n+1) U^*_{D-1}(1,n) \leq \frac{2\sigma}{e}.
\end{equation}

By setting $\widehat{\pi}_{D-1}(n) = \frac{1}{n+1}$ for each $n_{D-1} =n \in \{m-k(D-2),m-k(D-2)+1,\ldots, m\}$, as $k$ and $D$ are both finite, we obtain that $m \to \infty$ implies $n \to \infty$, and then obtain from Eqs.~\eqref{t-D}, \eqref{approx_1} and \eqref{squeeze} that
\begin{align}
     &\lim_{m \to \infty} (n+1)U_{D-1}\big((1,n),\frac{1}{n+1}\big) \notag \\
     & = \lim_{m \to \infty}  \big( \sigma (1-\frac{1}{n+1})^n + (n-k+1)U^*_{D}(1,n-k) \big) \notag \\
     &= \frac{2\sigma}{e}. \notag
\end{align}
Since $U^*_{D-1}(1,n)  \geq U_{D-1}\big((1,n),\frac{1}{n+1}\big)$, we have
\begin{align} \label{liminf}
    &\mathop{\lim\inf}_{m \to \infty} (n+1) U^*_{D-1}(1,n) \notag\\
    &\geq \mathop{\lim\inf}_{m \to \infty} (n+1)U_{D-1}\big((1,n),\frac{1}{n+1}\big) = \frac{2\sigma}{e}.
\end{align}
Combining inequalities~\eqref{limsup} and \eqref{liminf}, we have $\lim_{m \to \infty}(n+1) U^*_{D-1}(1,n) = \frac{2\sigma}{e}$.

For the case $t = D-2,D-3,\ldots,1$, iteratively repeating the above argument can lead to Eqs.~\eqref{lemma1} and \eqref{lemma12} for each possible $n_t = n$.

\section{Proof of Lemma 2}
As $U^*_{t}(1,0) = \sigma$ for each $t \in \mathcal{T}$, we have
\begin{equation} \label{ns0}
   U_t\big((1,1),p\big) = 2\sigma p (1-p) + (1-p)^2 U^*_{t+1}(1,1),
\end{equation}
for each  $t \in \mathcal{T} \setminus \{D\}$. Taking the derivative of $U_t\big((1,1),p\big)$ with respect to $p$ derives that
\begin{align}
    & \dfrac{\text{d}}{\text{d}p} U_t\big((1,1),p\big) \notag \\
    & = \big(2\sigma - 2U^*_{t+1}(1,1)\big) - \big( 4\sigma  - 2U^*_{t+1}(1,1) \big)p. \notag
\end{align}
As $\sigma > 0$ and $U^*_{t}(1,1) \leq \sigma$ for each $t \in \mathcal{T} \setminus \{D\}$,
we have
\begin{equation}\label{ns2}
\widehat{\pi}^*_t(1) = \frac{\sigma - U^*_{t+1}(1,1)}{2\sigma  - U^*_{t+1}(1,1)},
\end{equation}
for each  $t \in \mathcal{T} \setminus \{D\}$.
In particular,  as $U^*_D(1,1)=\sigma/4$, we obtain $\widehat{\pi}^*_{D-1}(1)=3/7$, which satisfies Eq.~\eqref{lemma22}.

Then, we aim to investigate the relation between $\widehat{\pi}^*_{t}(1)$ and $\widehat{\pi}^*_{D-1}(1)$ for each $t \in \mathcal{T} \setminus \{D-1,D\}$.
By setting $p=\widehat{\pi}^*_t(1)$ in Eq.~\eqref{ns0}, we obtain
\begin{align}
    U^*_t(1,1) & = 2\sigma \widehat{\pi}^*_t(1) \big( 1-\widehat{\pi}^*_t(1) \big) + \big( 1-\widehat{\pi}^*_t(1) \big)^2 U^*_{t+1}(1,1) \notag \\
    &= \dfrac{\sigma^2}{2\sigma - U^*_{t+1}(1,1)}. \label{ns3}
\end{align}
Using Eq.~\eqref{ns2} to express $U^*_{t+1}(1,1)$ and $U^*_{t}(1,1)$ in Eq.~\eqref{ns3} in terms of $\widehat{\pi}^*_{t}(1)$ and $\widehat{\pi}^*_{t-1}(1)$, respectively, we have
\begin{equation} \label{ns6}
    \widehat{\pi}^*_{t}(1) = \dfrac{\widehat{\pi}^*_{t+1}(1)}{1+\widehat{\pi}^*_{t+1}(1)},
\end{equation}
for each  $t \in \mathcal{T} \setminus \{D-1,D\}$.
Furthermore, recursively using Eq.~\eqref{ns6} yields
\begin{equation} \label{ns8}
    \widehat{\pi}^*_{t}(1) = \dfrac{ \widehat{\pi}^*_{D-1}(1) } { 1 + (D-t-1)\widehat{\pi}^*_{D-1}(1)}
\end{equation}
and thus implies Eq.~\eqref{lemma22} by $\widehat{\pi}^*_{D-1}(1)=3/7$.

Finally, combining Eqs.~\eqref{lemma22} and~\eqref{ns2} obtains
\begin{align}
    U^*_t(1,1) = \frac{1 - 2 \widehat{\pi}^*_{t-1}(1)}{1-\widehat{\pi}^*_{t-1}(1)}\sigma = \frac{3D-3t+1}{3D-3t+4}\sigma, \label{eq:u1}
\end{align}
for each $t \in \mathcal{T} \setminus \{1\}$, and substituting Eq.~\eqref{eq:u1} into Eq.~\eqref{ns3} obtains
$U^*_1(1,1) = \frac{3D-2}{3D+1}\sigma$.
Hence we complete the proof for Eq.~\eqref{lemma21}.

\section{Proof of Lemma 3}
We shall prove $U^{\text{eve}}_t(1,n) = \sigma \big(1-\frac{1}{D-t+1}\big)^n$ for each $n \in \mathcal{N}$ by induction from $t = D - 1$ down to 1.

First, when $t = D - 1$, by Eqs.~\eqref{trans},~\eqref{reward} and~\eqref{eve-bellman}, we have
\begin{align}
 & U^{\text{eve}}_{D-1}(1,n) \notag \\
 & = r_{D-1} \big((1,n),\widehat{\pi}_{D-1}^{\text{eve}}(n)\big) \notag \\
 & \quad + \sum_{n' \in \mathcal{N}} \beta_{D-1} \big((1,n'),(1,n),\widehat{\pi}_{D-1}^{\text{eve}}(n)\big) U^{\text{eve}}_{D}(1,n') \notag \\
 & = \sigma \frac{1}{2}\big(1 - \frac{1}{2}\big)^n + (1-\frac{1}{2})^{n+1} U^{\text{eve}}_{D}(1,0) \notag \\
 & = \sigma (1 - \frac{1}{2})^n, \notag
\end{align}
for each $n \in \mathcal{N}$, thereby establishing the induction basis.

Next, when $t \in \mathcal{T} \setminus \{D-1,D\}$, we assume $U^{\text{eve}}_{t+1}(1,n) = \sigma \big(1-\frac{1}{D-t}\big)^n$ for each $n \in \mathcal{N}$.
By Eqs.~\eqref{trans},~\eqref{reward} and~\eqref{eve-bellman}, we have
\begin{align}
 & U^{\text{eve}}_t(1,n) \notag \\
 & = r_t \big((1,n),\widehat{\pi}_t^{\text{eve}}(n)\big) \notag \\
 & \quad + \sum_{n' \in \mathcal{N}} \beta_t \big((1,n'),(1,n),\widehat{\pi}_t^{\text{eve}}(n)\big) U^{\text{eve}}_{t+1}(1,n') \notag \\
 & = \sigma \frac{1}{D-t+1}\big(1 - \frac{1}{D-t+1}\big)^n \notag \\
 & \quad + \sum_{n' \in \mathcal{N}} \binom{n}{n-n'} \big(\frac{1}{D-t+1}\big)^{n-n'} \notag \\
 & \quad\quad\quad\quad\,\, \cdot \big(1 - \frac{1}{D-t+1}\big)^{n'+1} \sigma \big(1-\frac{1}{D-t}\big)^{n'} \notag \\
 & = \sigma \big(1 - \frac{1}{D-t+1}\big)^n \frac{1}{D-t+1} \notag \\
 & \quad + \sigma \big(1 - \frac{1}{D-t+1}\big)^n \frac{D-t}{D-t+1} \notag \\
 & \quad\quad \cdot \sum_{n' \in \mathcal{N}} \binom{n}{n-n'} \big(\frac{1}{D-t}\big)^{n-n'} \big(1 - \frac{1}{D-t}\big)^{n'}\notag \\
 & = \sigma \big(1 - \frac{1}{D-t+1}\big)^n, \notag
\end{align}
for each $n \in \mathcal{N}$.
So, the inductive step is established.

Since both the base case and the inductive step have been proved as true, we have $U^{\text{eve}}_t(1,n) = \sigma \big(1-\frac{1}{D-t+1}\big)^n$ for each $t \in \mathcal{T} \setminus \{D\}$ and each $n \in \mathcal{N}$.

\section{Proof of Eq.~\eqref{heur_immd}}
Letting $f\big((M,\alpha),p\big) \triangleq \frac{(M + 1) \alpha}{\sigma} \sum_{n \in \mathcal{N}} b^{\text{bd}}(n) r_t\big((1,n),p\big)$ for $p \in [0,1]$ and $c_i = \binom{M+1}{i} \alpha^{i} (1-\alpha)^{(M+1-i)}$ for $1 \leq i \leq M + 1$, we have
\begin{align}
    %f\big((M,\alpha),p\big) &\triangleq \frac{(M + 1) \alpha}{\sigma} \sum_{n \in \mathcal{N}} b^{\text{bd}}(n) r_t\big((1,n),p\big) \notag \\
    f\big((M,\alpha),p\big) = \sum_{i = 1}^{M + 1} i c_i p(1-p)^{i-1}. \notag
\end{align}
The derivative of $f\big((M,\alpha),p\big)$ with respect to $p$ is given by
\begin{align}
    & \dfrac{\text{d}}{\text{d}p} f\big((M,\alpha),p\big) \notag \\
    & = \sum_{i=1}^{M + 1} i c_i (1-p)^{i-1} - \sum_{i=2}^{M + 1} i(i-1) c_i p(1-p)^{i-2} \notag \\
    & = (M+1)\alpha + (M+1)^2 \alpha^{M+1} (-p)^{M} + \sum_{j=1}^{M-1} \beta_j p^j, \label{derivative}
\end{align}
where $\beta_j$, $1 \leq j \leq M - 1$ is derived as follows:
\begin{align}
    \beta_j & = (-1)^j \sum_{k=1}^{M + 1 - j} \binom{M+1}{j+k} \alpha^{j+k} (1-\alpha)^{M+1-j-k}  \notag \\
    & \quad \cdot (j+k) \bigg[\binom{j+k-1}{j} + (j+k-1)\binom{j+k-2}{j-1}\bigg] \notag \\
    & = (-1)^j (j+1)^2 \alpha^{j+1}  \notag \\
    & \quad \cdot \sum_{k=1}^{M + 1 - j} \binom{j+k}{k-1} \binom{M+1}{j+k} \alpha^{k-1} (1-\alpha)^{M+1-j-k} \notag \\
    & = (-1)^j (j+1)^2 \alpha^{j+1} \binom{M + 1}{j+1} \notag \\
    & \quad \cdot \sum_{k=1}^{M + 1 -j} \binom{M - j}{k-1} \alpha^{k-1} (1-\alpha)^{M-j - k+1} \notag \\
    & = (-1)^j \alpha^{j+1} \notag \\
    & \quad \cdot \big( j(M+1)^2 + (M+1)(M-j) \big) \frac{(M-1)!}{(M-j)!j!} \notag \\
    & = (-1)^j \alpha^{j+1} \notag \\
    & \quad \cdot \bigg[ (M+1)^2 \binom{M-1}{j-1} + (M+1)\binom{M-1}{j} \bigg]. \label{j3}
\end{align}

Combining Eqs.~\eqref{derivative} and~\eqref{j3}, we have
\begin{align}
    & \dfrac{\text{d}}{\text{d}p} f\big((M,\alpha),p\big) \notag \\
    & = (M+1) \alpha + (M+1)^2 \alpha^{M+1} (-p)^{M} \notag \\
    & \quad + \sum_{j=1}^{M-1} \Big[ (M+1)^2 \binom{M-1}{j-1} + (M+1) \binom{M-1}{j} \Big] \notag \\
    & \quad\quad\quad \cdot \alpha^{j+1} (-p)^j \notag \\
    & =  (M+1) \alpha \big( 1 - (M+1)\alpha p \big)  \notag \\
    & \quad + (M+1) \alpha \big( 1 - (M+1) \alpha p \big) (-\alpha p)^{M-1} \notag \\
    & \quad + \sum_{j=1}^{M-2} \binom{M-1}{j} (M+1) \alpha \big( 1 - (M+1) \alpha p \big) (-\alpha p)^{j} \notag \\
    %& = (M+1) \alpha \big( 1 - (M+1) \alpha p \big) \notag \\
    %& \quad \cdot \big( 1 + \sum_{j=1}^{M-2} \binom{M-1}{j} (-\alpha p)^{j} + (-\alpha p)^{M-1} \big) \notag \\
    & = (M+1) \alpha \big( 1 - (M+1) \alpha p \big) \big( 1 - \alpha p \big)^{M-1}. \label{simple_deri}
\end{align}
From Eq.~\eqref{simple_deri}, for $p \in [0,1]$, we obtain that $f\big((M,\alpha),p\big) \leq f\big((M,\alpha),\frac{1}{M\alpha + \alpha}\big)$ when $\frac{1}{M\alpha + \alpha} \leq 1$, and $f\big((M,\alpha),p\big) \leq f\big((M,\alpha),1\big)$ when $\frac{1}{M\alpha + \alpha} > 1$.

Hence we complete the proof for Eq.~\eqref{heur_immd}.

\section*{Acknowledgement}
The authors would like to thank Dr. He Chen for helpful suggestions and discussions.

\bibliographystyle{IEEEtran}
\bibliography{pomdpBroadcast}

\begin{thebibliography}{10}
\providecommand{\url}[1]{#1}
\csname url@samestyle\endcsname
\providecommand{\newblock}{\relax}
\providecommand{\bibinfo}[2]{#2}
\providecommand{\BIBentrySTDinterwordspacing}{\spaceskip=0pt\relax}
\providecommand{\BIBentryALTinterwordstretchfactor}{4}
\providecommand{\BIBentryALTinterwordspacing}{\spaceskip=\fontdimen2\font plus
\BIBentryALTinterwordstretchfactor\fontdimen3\font minus
  \fontdimen4\font\relax}
\providecommand{\BIBforeignlanguage}[2]{{%
\expandafter\ifx\csname l@#1\endcsname\relax
\typeout{** WARNING: IEEEtran.bst: No hyphenation pattern has been}%
\typeout{** loaded for the language `#1'. Using the pattern for}%
\typeout{** the default language instead.}%
\else
\language=\csname l@#1\endcsname
\fi
#2}}
\providecommand{\BIBdecl}{\relax}
\BIBdecl

\bibitem{2019urllc}
D.~{Feng}, C.~{She}, K.~{Ying}, L.~{Lai}, Z.~{Hou}, T.~Q.~S. {Quek}, Y.~{Li},
  and B.~{Vucetic}, ``Toward ultrareliable low-latency communications: Typical
  scenarios, possible solutions, and open issues,'' \emph{IEEE Veh. Technol.
  Mag.}, vol.~14, no.~2, pp. 94--102, 2019.

\bibitem{2018Gao}
J.~Gao, M.~Li, L.~Zhao, and X.~Shen, ``Contention intensity based distributed
  coordination for {V2V} safety message broadcast,'' \emph{IEEE Trans. Veh.
  Technol.}, vol.~67, no.~12, pp. 12\,288--12\,301, 2018.

\bibitem{2019Luvi}
M.~Luvisotto, Z.~Pang, and D.~Dzung, ``High-performance wireless networks for
  industrial control applications: New targets and feasibility,'' \emph{Proc.
  IEEE}, vol. 107, no.~6, pp. 1074--1093, 2019.

\bibitem{2013CRN}
Y.~H. {Bae}, ``Analysis of optimal random access for broadcasting with deadline
  in cognitive radio networks,'' \emph{IEEE Commun. Lett.}, vol.~17, no.~3, pp.
  573--575, 2013.

\bibitem{2013RA}
Y.~H. {Bae}, ``Random access scheme to improve broadcast reliability,''
  \emph{IEEE Commun. Lett.}, vol.~17, no.~7, pp. 1467--1470, 2013.

\bibitem{2015Queue}
Y.~H. {Bae}, ``Queueing analysis of deadline-constrained broadcasting in
  wireless networks,'' \emph{IEEE Commun. Lett.}, vol.~19, no.~10, pp.
  1782--1785, 2015.

\bibitem{2011WAVE}
C.~{Campolo}, A.~{Vinel}, A.~{Molinaro}, and Y.~{Koucheryavy}, ``Modeling
  broadcasting in {IEEE} 802.11p/{WAVE} vehicular networks,'' \emph{IEEE
  Commun. Lett.}, vol.~15, no.~2, pp. 199--201, 2011.

\bibitem{2018asymAloha}
L.~{Deng}, J.~{Deng}, P.~{Chen}, and Y.~S. {Han}, ``On the asymptotic
  performance of delay-constrained slotted {ALOHA},'' in \emph{Proc. IEEE
  ICCCN}, 2018, pp. 1--8.

\bibitem{2019mprAloha}
Y.~{Zhang}, Y.~{Lo}, F.~{Shu}, and J.~{Li}, ``Achieving maximum reliability in
  deadline-constrained random access with multiple-packet reception,''
  \emph{IEEE Trans. Veh. Technol.}, vol.~68, no.~6, pp. 5997--6008, 2019.

\bibitem{2020TTS}
L.~{Deng}, F.~{Liu}, Y.~{Zhang}, and W.~S. {Wong}, ``Delay-constrained
  topology-transparent distributed scheduling for {MANETs},'' \emph{IEEE Trans.
  Veh. Technol.}, vol.~70, no.~1, pp. 1083--1088, 2021.

\bibitem{Segall1976}
A.~{Segall}, ``Recursive estimation from discrete-time point processes,''
  \emph{IEEE Trans. Inf. Theory}, vol.~22, no.~4, pp. 422--431, 1976.

\bibitem{1987Bayesbroadcast}
R.~{Rivest}, ``Network control by {Bayesian} broadcast,'' \emph{IEEE Trans.
  Inf. Theory}, vol. IT-33, no.~3, pp. 323--328, 1987.

\bibitem{2004fine}
G.~{del Angel} and T.~L. {Fine}, ``Optimal power and retransmission control
  policies for random access systems,'' \emph{IEEE/ACM Trans. Netw.}, vol.~12,
  no.~6, pp. 1156--1166, 2004.

\bibitem{2004Dynamic802.11}
L.~{Bononi}, M.~{Conti}, and E.~{Gregori}, ``Runtime optimization of {IEEE}
  802.11 wireless {LAN}s performance,'' \emph{IEEE Trans. Parallel Distrib.
  Syst.}, vol.~15, no.~1, pp. 66--80, 2004.

\bibitem{2013FASA}
H.~{Wu}, C.~{Zhu}, R.~J. {La}, X.~{Liu}, and Y.~{Zhang}, ``{FASA}: Accelerated
  {S-ALOHA} using access history for event-driven {M2M} communications,''
  \emph{IEEE/ACM Trans. Netw.}, vol.~21, no.~6, pp. 1904--1917, 2013.

\bibitem{1971POMDP}
R.~{Smallwood} and E.~{Sondik}, ``The optimal control of partially observable
  {Markov} processes over a finite horizon,'' \emph{Oper. Res.}, vol.~21,
  no.~5, pp. 1071--1088, 1973.

\bibitem{2019pcsma}
Y.~{Zhang}, A.~{Gong}, Y.~{Lo}, J.~{Li}, F.~{Shu}, and W.~S. {Wong},
  ``Generalized $p$-persistent {CSMA} for asynchronous multiple-packet
  reception,'' \emph{IEEE Trans. Commun.}, vol.~67, no.~10, pp. 6966--6979,
  2019.

\bibitem{2018energy}
A.~{Biason}, S.~{Dey}, and M.~{Zorzi}, ``A decentralized optimization framework
  for energy harvesting devices,'' \emph{IEEE Trans. Mob. Comput.}, vol.~17,
  no.~11, pp. 2483--2496, 2018.

\bibitem{2019traffic}
\emph{A {5G} traffic model for industrial use cases}.\hskip 1em plus 0.5em
  minus 0.4em\relax White Paper, 5G Alliance for Connected Industries and
  Automation, 2019.

\bibitem{pm2014}
M.~L. {Puterman}, \emph{Markov decision processes: Discrete stochastic dynamic
  programming}.\hskip 1em plus 0.5em minus 0.4em\relax John Wiley \& Sons,
  2014.

\bibitem{kumar2015}
P.~R. {Kumar} and P.~{Varaiya}, \emph{Stochastic systems: Estimation,
  identification, and adaptive control}.\hskip 1em plus 0.5em minus 0.4em\relax
  SIAM, 2015.

\bibitem{bernstein2002complexity}
D.~S. {Bernstein}, R.~Givan, N.~Immerman, and S.~Zilberstein, ``The complexity
  of decentralized control of {Markov} decision processes,'' \emph{Math. Oper.
  Res.}, vol.~27, no.~4, pp. 819--840, 2002.

\end{thebibliography}

%\end{IEEEbiography}


%\newpage


\end{document}